\documentclass[letter]{aa}     

\usepackage{graphicx}
\usepackage{txfonts}
\usepackage{lipsum}
\usepackage{subcaption}         
\usepackage{lscape}             
\usepackage{placeins}           
                                
\usepackage{xcolor}
\usepackage{comment}

\begin{document}

   \title{A high-frequency turnover in the persistent radio source of FRB\,20201124A}


%
%
%

   \author{G. Bruni\inst{1}\corrauth{gabriele.bruni@inaf.it}        
        \and G. Gianfagna\inst{1}
        \and L. Piro\inst{1}
        \and A.L. Thakur\inst{1}
        \and L. Nicastro\inst{2}
        \and E. Palazzi\inst{2}
        \and A. Rossi\inst{2}
        \and S. Savaglio\inst{3,2,4}
        \and A. Gardini\inst{5}
        \and Y.-P. Yang\inst{6}
        \and B. Zhang\inst{7,8,9,10}
        }

   \institute{INAF -- Istituto di Astrofisica e Planetologia Spaziali, Via del Fosso del Cavaliere 100, Roma, 00133, Italy\\
             \email{gabriele.bruni@inaf.it}
    \and
    INAF -- Osservatorio di Astrofisica e Scienza dello Spazio di Bologna, via Piero Gobetti 93/3, I-40129 Bologna, Italy
    \and
    Dipartimento di Fisica, Università della Calabria, Arcavacata di Rende, Italy 
    \and
    Laboratori Nazionali di Frascati, INFN (Istituto Nazionale di Fisica Nucleare), Frascati, Italy 
    \and
    Instituto de Astrofísica de Andalucía, CSIC (Consejo Superior de Investigaciones Científicas), Granada, Spain
    \and
    South-Western Institute for Astronomy Research, Yunnan Key Laboratory of Survey Science, Yunnan University, Kunming, Yunnan 650504, People's Republic of China 
    \and
    The Hong Kong Institute for Astronomy and Astrophysics, the University of Hong Kong, Pokfulam Road, Hong Kong
    \and
    Department of Physics, the University of Hong Kong, Pokfulam Road, Hong Kong
    \and
    Nevada Center for Astrophysics, University of Nevada, Las Vegas, NV, USA. 
    \and
    Department of Physics and Astronomy, University of Nevada, Las Vegas, NV, USA
   }

   \date{Received September 30, 20XX}

 \abstract
{Persistent radio sources (PRSs) associated with repeating fast radio bursts probe the immediate magneto-ionic environment of their engine and thereby constrain the nature and evolutionary stage of the progenitor. The persistent radio source of FRB\,20201124A is the most extreme in the current sample, with a spectrum that remains flat or inverted up to 22\,GHz.}
{We aim to constrain the centimetre-to-millimetre spectral energy distribution of this source and to test whether its properties are consistent with a young, rotation-powered magnetar wind nebula.}
{We analysed archival ALMA Band\,3 observations at $\sim100$\,GHz and combined them with existing VLA measurements at 6, 15 and 22\,GHz. The resulting spectrum was fitted with the Granot--Sari broken power law in the slow-cooling regime using Bayesian nested sampling.}
{The two-parameter model yields a peak frequency $\nu_m = 63^{+100}_{-33}$\,GHz and a peak flux $F_m = 28\pm9\,\mu$Jy. This solution is consistent with theoretical expectations for a young magnetar wind nebula, which predict a peak frequency in the same range and a self-absorption frequency below $\sim0.5$\,GHz.}
{With its inverted spectrum extending to tens of gigahertz and the highest peak frequency among known PRS, the source associated with FRB\,20201124A is the most extreme in the current sample. Its properties match those expected for a young ($t\lesssim20$\,yr), rotation-powered magnetar wind nebula, making it a strong candidate for the youngest magnetar associated with a repeating FRB observed to date.}


   \maketitle
\nolinenumbers

\section{Introduction}
Fast radio bursts (FRBs) are millisecond-duration radio transients of extragalactic origin, with dispersion measures (column density of free electrons) that place the majority of them at cosmological distances \citep{Petroff2019,Cordes2019}. A growing fraction of FRBs are observed to repeat, and the repeating population has enabled precise localisations and detailed studies of their host galaxies and local environments \citep{Spitler2016,CHIME2021}. A subset of repeating FRBs is associated with compact PRSs, whose luminosities and spectra are difficult to reconcile with star formation alone and have been interpreted as emission from a magnetar wind nebula or a young supernova remnant surrounding the central engine \citep{Marcote2017,Margalit2018,Bruni2024Nat,Bruni2025AandA,2026ApJ...996L..16M,Bruni2026,Pelliciari2026}.


\begin{figure*}
    \centering
    \includegraphics[width=1.0\columnwidth]{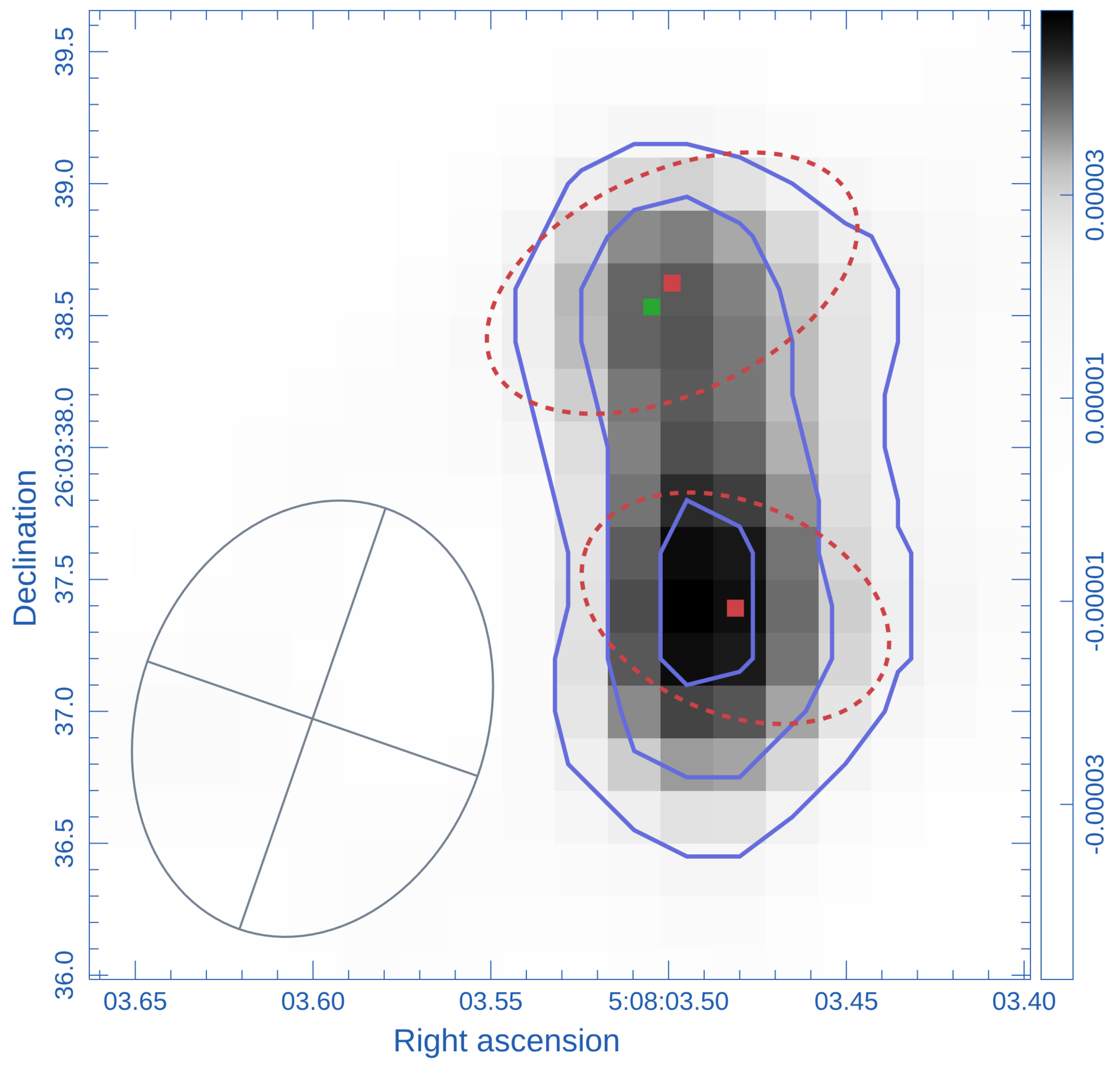}
    \includegraphics[width=1.0\columnwidth]{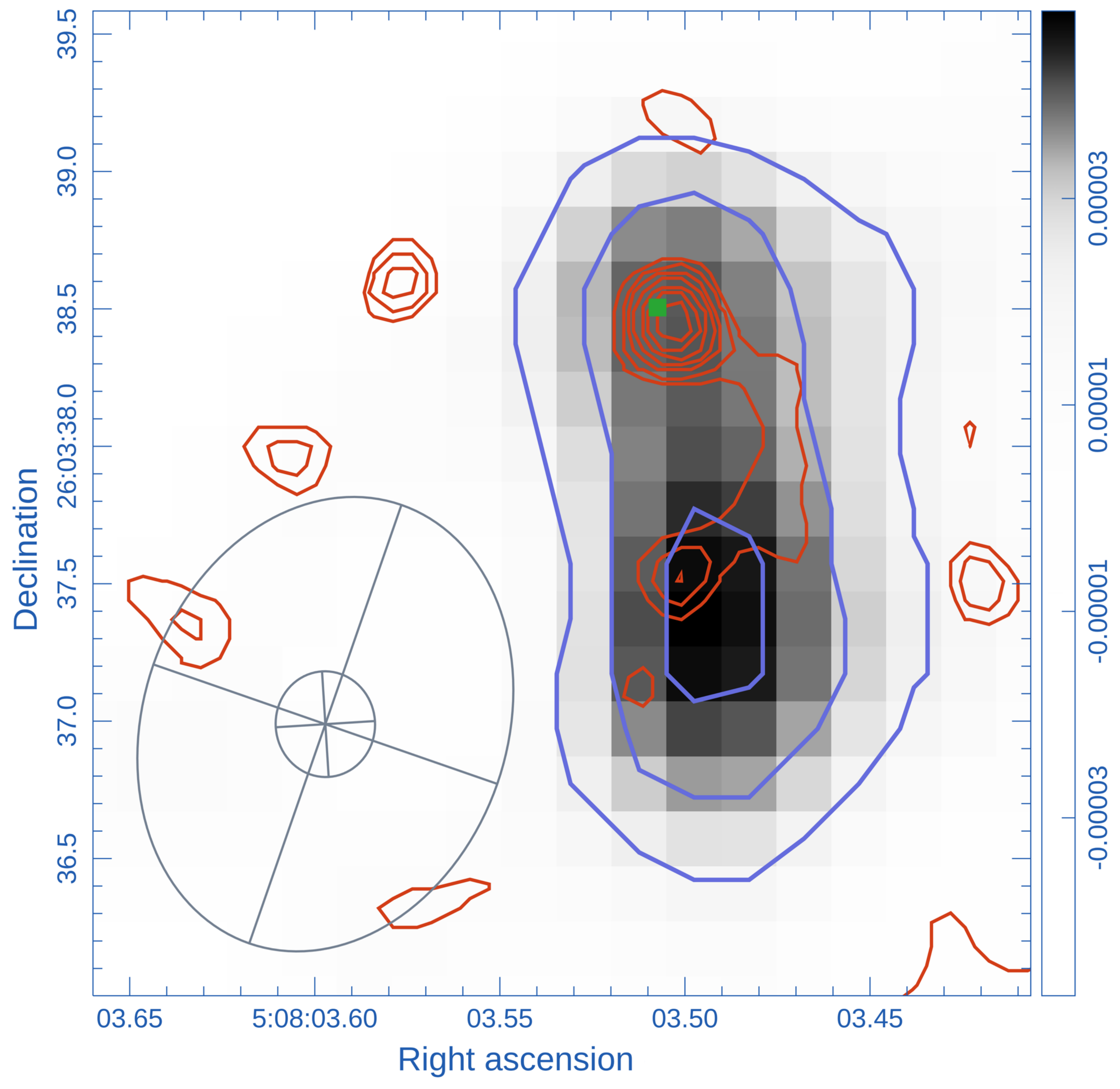}
    \caption{Left: ALMA Band~3 image of the host of FRB\,20201124A at $\sim100$\,GHz (colour scale, in Jy\,beam$^{-1}$). The emission was fitted with two Gaussian components (dashed ellipses indicate the major and minor axes FWHM) centred on the positions marked by the red squares. Contours are 2, 3, 4$\times$ the local rms of 11\,$\mu$Jy\,beam$^{-1}$. The green square indicates the milliarcsecond FRB position reported by \citet{Nimmo_2022}. Right: the same ALMA image with superimposed VLA 15\,GHz contours from \citet{Bruni2024Nat} (red; 3--9$\times$ the local rms of 2.3\,$\mu$Jy\,beam$^{-1}$). The northern ALMA peak and the compact VLA persistent source coincide with the FRB within the positional uncertainties; the southern ALMA peak traces the host centre. The synthesised ALMA beam ($1.69''\times1.33''$) is shown in the lower left corner of both panels, together with the VLA beam in the right panel ($0.38''\times0.36''$, small circle).}
    \label{fig:ALMA+VLA}
\end{figure*}


PRSs are the immediate magneto-ionic environment of the FRB central engine. Their luminosities and rotation measures therefore provide direct constraints on the progenitor and on the evolutionary stage of the surrounding nebula. A simple relation between PRS radio luminosity and FRB Faraday rotation measure (RM) was predicted on theoretical grounds by \citet{yang20,Yang2022}, under the assumption that both quantities arise from the same magnetised plasma. This relation was observationally established and extended into the low-luminosity, low-RM regime by the detection of the PRS associated with FRB\,20201124A \citep{Bruni2024Nat}. More recently, \citet{Yang2026} showed that the intrinsic scatter of the $L_\nu$--$|{\rm RM}|$ relation encodes information on the temporal evolution of the nebula radius, offering a diagnostic of the expansion history of the FRB environment.

The repeating fast radio burst FRB\,20201124A was localised to a barred spiral galaxy at $z=0.0979$ and found to be embedded in a region of elevated star formation \citep{Piro2021,Fong2021}. Multi-frequency VLA observations revealed a compact persistent radio source coincident with the milliarcsecond FRB position, with a spectrum that remains flat or inverted up to 22\,GHz \citep{Bruni2024Nat}. Although high-resolution VLBI observations have not yet yielded a detection (see Sect.~\ref{sec:hsa}), the source is considered a bona fide PRS on the basis of several independent indicators: (i) its precise positional coincidence with the FRB, (ii) its inverted spectrum, which is inconsistent with the steep or flat spectra typical of star-formation-related emission, and (iii) the fact that the star-formation rate required to account for its radio luminosity ($\gtrsim 3\,M_\odot\,\mathrm{yr}^{-1}$) exceeds both the H$\alpha$-derived SFR measured with GTC/MEGARA at the FRB location ($\sim0.4\,M_\odot\,\mathrm{yr}^{-1}$) and the upper limit from NOEMA millimetre observations ($<2.2\,M_\odot\,\mathrm{yr}^{-1}$) \citep{Bruni2024Nat}. This inverted spectrum already set FRB\,20201124A apart from the other known PRSs, whose emission typically peaks at lower frequencies and declines above a few gigahertz, and made it the most extreme object in the current sample in terms of spectral energy distribution.

First indications of the spectral shape of other PRSs have recently become available. \citet{Bruno2026} reported a flat spectrum ($\alpha = 0.20\pm0.05$) between 1 and 6\,GHz for the PRS associated with FRB\,20190417A, and placed a lower limit on the turnover frequency ($\nu_{\rm to} > 370$\,MHz at 95\,\% confidence) from the non-detection at 144\,MHz. \citet{Pelliciari2026} presented a similarly tentative SED for another PRS, constructed from a single VLBI measurement and two upper limits. While these data already hint at a possible turnover at relatively low frequencies, the limited sampling and mixed resolution make the constraints preliminary.

Subsequent modelling of PRS spectra has interpreted the radio emission as synchrotron radiation from a magnetar wind nebula or a young supernova remnant, with the peak frequency marking the transition between optically thick and thin regimes \citep{Rahaman2025}. In the Granot--Sari formalism adopted by \citet{Rahaman2025}, the location of the peak and the spectral indices on either side constrain the magnetic field, the electron energy distribution, and the size of the emitting region. For FRB\,20201124A the existing data do not go beyond 22\,GHz, leaving the peak frequency and the optically thin slope essentially unconstrained.

In this work we present the first radio-to-millimetre spectral energy distribution of the PRS associated with FRB\,20201124A, extending from 6 to $\sim100$\,GHz with new ALMA Band\,3 observations. We fit the SED with the Granot--Sari model as formulated by \citet{Rahaman2025} and derive the first quantitative constraints on the physical parameters of the emitting region. Throughout the
work we adopt a standard $\Lambda$CDM cosmology \citep{Planck18} with $\rm H_0 =  67.4 \rm \ km \ s^{-1}Mpc^{-1}$, $\Omega_m = 0.315$ and $\Omega_{\Lambda} = 0.685$.


\begin{figure*}
  \centering
  \includegraphics[width=1.2\columnwidth]{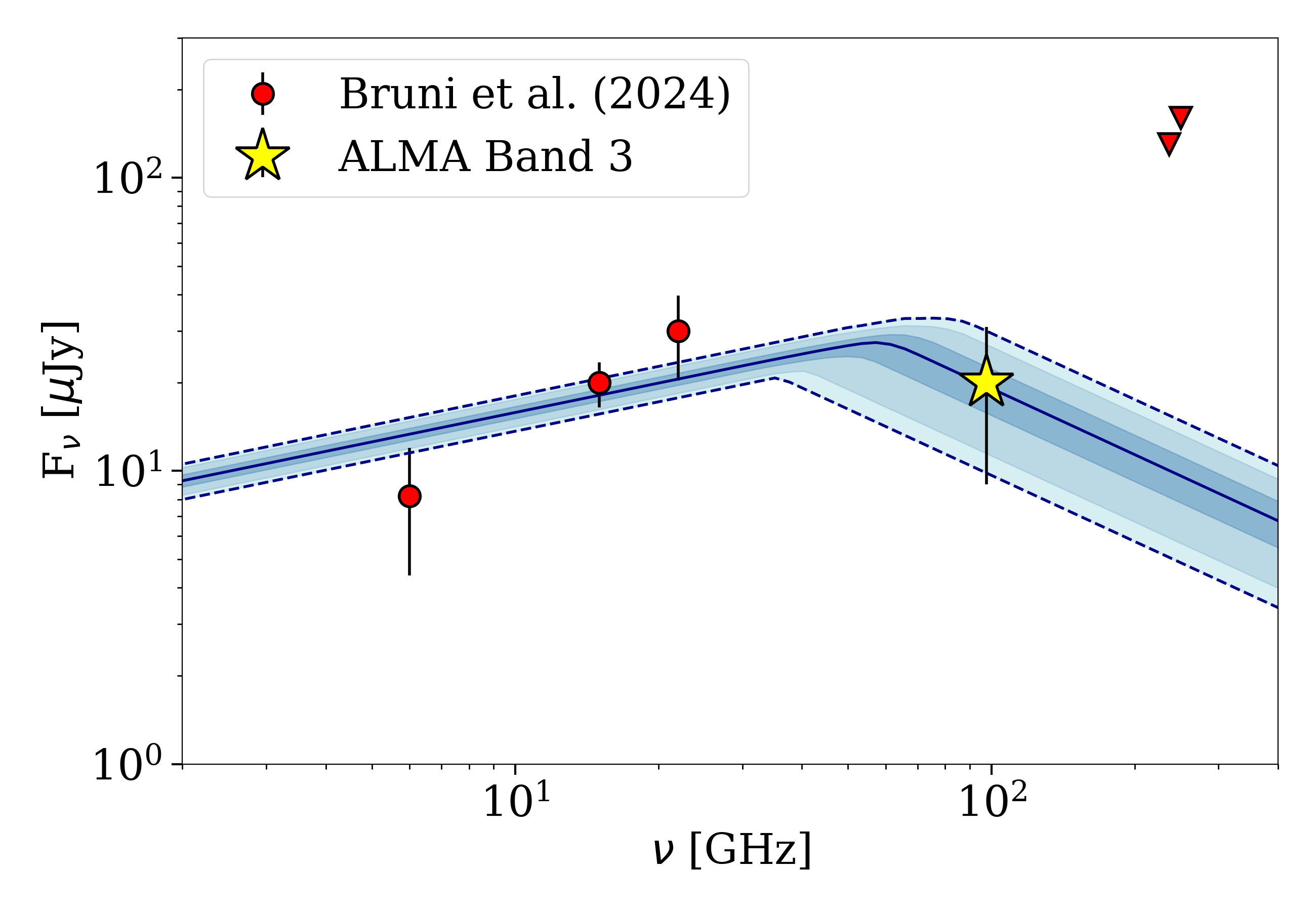}
  \hfill
  \includegraphics[width=0.8\columnwidth]{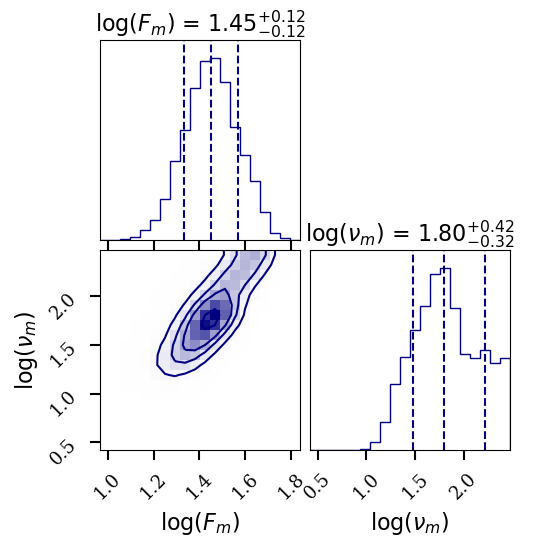}
  \caption{Left panel: radio-to-millimetre spectrum of the persistent source associated with FRB\,20201124A. Red points are VLA measurements at 6, 15 and 22\,GHz from \citet{Bruni2024Nat}; the yellow star is the ALMA Band\,3 detection at $\sim$100\,GHz presented in this work. Upper limits from NOEMA \citep{Bruni2024Nat} are shown as downward arrows. The solid line is the median of the 10\,000 best-likelihood spectra, and the shaded blue areas within the dashed lines show the 1, 2 and 3$\sigma$ credible regions. Right panel: corner plot of the two-parameter fit shown in the left panel, made using {\tt{corner}} \citep{corner}.}
  \label{fig:prs_spectrum}
\end{figure*}


\section{Observations and archival data}

\subsection{ALMA}
\label{sec:alma}

We used archival ALMA Band~3 observations from project
2023.1.01090.S (PI:~Hatsukade), taken on 2023 December~31.
The data are centred at $\sim97$\,GHz, with a synthesised beam of
$1.69''\times1.33''$ at a position angle of $-19^\circ$ in the
12-m array configuration. The epoch is $\sim2$\,yr after the VLA
observations reported by \citet{Bruni2024Nat}.

The measurement set was calibrated with the ALMA pipeline and
imaged in CASA \citep{2007ASPC..376..127M} with natural weighting, chosen to minimise the
noise. The rms in the final image is $11\,\mu$Jy\,beam$^{-1}$.
The host is detected and appears elongated north--south, tracing
the bar seen in the HST imaging of \citet{Bruni2024Nat},
with a brightness peak near the optical centre of the galaxy.

Because the synthesised beam is larger than the projected separation between the FRB and the host-galaxy centre, the Gaussian wings of the point-spread function of the brighter central component contaminate the measurement at the FRB position. To separate the contribution of the central component from that at the FRB position, we fitted the elongated structure with two Gaussians using the CASA task \texttt{imfit}, with initial estimates taken from a preliminary two-component fit and providing the image rms so that the fit properly accounts for the noise. The fit
converged to a northern component at
$05^{\mathrm{h}}08^{\mathrm{m}}03.5016^{\mathrm{s}}$,
$+26^{\circ}03'38.595''$
and a southern component at
$05^{\mathrm{h}}08^{\mathrm{m}}03.483^{\mathrm{s}}$,
$+26^{\circ}03'37.363''$ (see Fig. \ref{fig:ALMA+VLA}, left panel).
The northern position is consistent with the milliarcsecond FRB
localisation of \citet{Nimmo_2022}; the southern position coincides
with the host centre.
The integrated flux densities returned by the two-component fit are $20\pm11\,\mu$Jy for the northern component and $46\pm19\,\mu$Jy for the southern one. The uncertainties combine the image noise with the formal error of the Gaussian fit. We note that the emission at the FRB position is significant at $>3\sigma$ in the image before deblending; the lower flux density obtained for the northern component is the expected result of subtracting the contribution from the Gaussian wings of the brighter southern component. At this signal-to-noise ratio both components remain unresolved, so the reported values should be interpreted as point-source flux densities.

Figure~\ref{fig:ALMA+VLA} (right panel) overlays the ALMA contours (black) on the
VLA 15\,GHz contours (red) from \citet{Bruni2024Nat}. The FRB
position, the compact persistent source identified with the VLA,
and the northern ALMA component agree within the positional
uncertainties.

\subsection{HSA}
\label{sec:hsa}

We additionally observed the PRS of FRB\,20201124A with the High
Sensitivity Array (HSA) in two epochs, both centered to the
milliarcsecond FRB position of \citet{Nimmo_2022}. The first epoch
(BB464B) was taken on 2025 April 20--21, from 21:31 to 02:48\,UTC
($\sim5.3$\,h on source), with eleven stations: the VLBA antennas
BR, FD, HN, KP, LA, MK, NL, OV, PT and SC, plus the phased VLA (Y).
The second epoch (BB464A) was taken on 2025 May 16, from 13:45 to
19:03\,UTC (also $\sim5.3$\,h), with ten stations: BR, HN, KP,
LA, MK, NL, OV, PT, plus Y and Effelsberg (EB). Both epochs used four 128\,MHz spectral
windows centred at 15.176, 15.304, 15.432 and 15.560\,GHz (total
bandwidth 512\,MHz), recording full polarisation. The data were
calibrated with the rPICARD pipeline \citep{Janssen2019}, and imaged in CASA.  
With natural weighting the image rms is $8\,\mu$Jy\,beam$^{-1}$. No source is detected at the FRB position, corresponding to a $5\sigma$ upper limit of $40\,\mu$Jy\,beam$^{-1}$. This limit lies above the VLA flux density of $20.0\pm3.5\,\mu$Jy measured at the same frequency by \citet{Bruni2024Nat}, so the observation is not deep enough to test whether the persistent emission is compact on parsec scales. A detection, or a limit below the VLA flux, would require a significantly lower rms. The non-detection is therefore not in tension with the VLA flux density and does not challenge the association of the persistent source with the FRB.


\section{Modelling the magnetar wind nebula spectrum}
\label{sec:prs_spectrum}

The centimetre-to-millimetre spectrum of the persistent source associated with FRB\,20201124A is not a single power law. VLA measurements at 6, 15 and 22\,GHz give $8.2\pm3.8$, $20.0\pm3.5$ and $30.0\pm9.7\,\mu$Jy \citep{Bruni2024Nat}, while our ALMA detection at $\sim100$\,GHz, $20\pm11\,\mu$Jy, lies below a continuation of the rising VLA spectrum. The spectral turnover suggested by these four measurements is the key observable that we compare with the predictions of a magnetar-wind nebula (MWN) model.

We adopt the slow-cooling synchrotron spectrum of \citet{Rahaman2025}, which is the Granot--Sari broken power law \citep{Granot2002} in the slow-cooling phase ($\nu_{\rm sa}<\nu_m<\nu_c$). Synchrotron self-absorption is predicted below $\nu_{\rm sa}\sim0.5$\,GHz for the nebular parameters of this source, so it does not shape the VLA--ALMA window. Between $\nu_{\rm sa}$ and the injection break $\nu_m$ the optically thin spectrum rises as
\begin{equation}
  F_\nu \propto \nu^{1/3},
  \label{eq:low}
\end{equation}
while above $\nu_m$ and below the cooling break it falls as
\begin{equation}
  F_\nu \propto \nu^{-(p-1)/2}.
  \label{eq:high}
\end{equation}
We fix the electron index $p$ at the value used by \citet{Rahaman2025}, $p=2.5$, which gives a high-frequency slope of $-0.75$. The only free parameters are then the break frequency $\nu_m$ and the flux density $F_m$ at that break. The model spectrum is normalised so that $F_m$ is the flux of the two asymptotic branches at $\nu_m$; the smooth join peaks slightly below $\nu_m$.

We fit the radio spectrum with this broken power-law model using Bayesian statistics. In particular, we use a dynamic nested sampling algorithm through the {\tt{dynesty}} python package \citep{Higson2019, Speagle2020, koposov2024}. The two parameters have log-uniform priors: $\nu_m$ in 1--300\,GHz and peak flux $F_m$ between 1 and 500\,$\mu$Jy. We find
\begin{equation}
  \nu_m = 63^{+100}_{-33}\,\mathrm{GHz}, \qquad
  F_m = 28\pm9\,\mu\mathrm{Jy}.
\end{equation}
See Fig.~\ref{fig:prs_spectrum}. The spectral peak luminosity corresponding to $\nu_m$ is $L_m = 6.7 \times 10^{27}\,\mathrm{erg\,s^{-1}\,Hz^{-1}}$.

This model fits the data well, but slightly over-predicts the flux at 6\,GHz. For completeness we also performed a three-parameter fit that includes the self-absorption frequency $\nu_{\rm sa}$ as a free parameter. However, this more complex model is not statistically preferred. The Akaike Information Criterion favours the simpler two-parameter model, and the Bayes factor likewise supports the baseline model. We therefore adopt the two-parameter solution as our primary result. This is also consistent with the theoretical expectations of \citet{Rahaman2025}, who predict that the self-absorption frequency of a magnetar wind nebula should lie below $\sim 0.5$\,GHz. Such a low value does not affect the detectability of the FRB bursts, which have been observed down to a few hundred MHz.


\section{Conclusions}
The centimetre-to-millimetre spectrum of the persistent source associated with FRB\,20201124A shows a clear turnover. Bayesian fits with the Granot--Sari broken power law of \citet{Rahaman2025} (slow-cooling regime, $p=2.5$) yield a peak frequency $\nu_m = 63^{+100}_{-33}\,\mathrm{GHz}$ and a peak flux $F_m = 28\pm9\,\mu\mathrm{Jy}$. The model is consistent with the theoretical expectations of \citet{Rahaman2025}, who predict a self-absorption frequency below $\sim 0.5$\,GHz for a young magnetar wind nebula. At such frequencies the nebula remains optically thin to the FRB emission, in agreement with the detection of bursts down to a few hundred MHz.

In the young-magnetar scenario of \citet{Rahaman2025}, a rotation-powered magnetar wind nebula can sustain the observed radio luminosity only while the system is still young ($t\lesssim20\,\mathrm{yr}$). Beyond this age the spin-down luminosity drops too rapidly, and only internal magnetic-field decay can power the brighter PRSs associated with other repeating FRBs. The relatively low luminosity of the PRS of FRB\,20201124A is therefore naturally explained if the nebula is rotation-dominated and only a few tens of years old. The measured peak frequency ($\nu_m\sim50$--$60\,\mathrm{GHz}$) and the absence of a detectable self-absorption break in the observed band match the expectations of \citet{Rahaman2025} for a compact, still-young nebula whose expansion has been slowed by a relatively massive, low-energy ejecta shell.

With its inverted spectrum extending to tens of gigahertz and the highest peak frequency among known PRSs, the persistent source associated with FRB\,20201124A is the most extreme in the current sample in terms of spectral properties. This makes it a strong candidate for the youngest magnetar associated with a repeating FRB observed to date. Concluding, an inverted spectrum extending to tens of gigahertz and a self-absorption break well below the peak are the characteristic signatures of a young, rotation-powered magnetar wind nebula, precisely the regime favoured by our data.


\begin{acknowledgements}
The research leading to these results has received funding from the European Union’s Horizon 2020 programme under the AHEAD2020 project (grant agreement no. 871158).
Y.P.Y is supported by the National Natural Science Foundation of China (No. 12473047), the National Key Research and Development Program of China (No. 2024YFA1611603) and the Yunnan Key Laboratory of Survey Science (No. 202449CE340002).
This paper makes use of the following ALMA data: ADS/JAO.ALMA\#2023.1.01090.S. ALMA is a partnership of ESO (representing its member states), NSF (USA) and NINS (Japan), together with NRC (Canada), NSTC and ASIAA (Taiwan), and KASI (Republic of Korea), in cooperation with the Republic of Chile. The Joint ALMA Observatory is operated by ESO, AUI/NRAO and NAOJ.
The National Radio Astronomy Observatory and Green Bank Observatory are facilities of the U.S. National Science Foundation operated under cooperative agreement by Associated Universities, Inc.
Partly based on observations with the 100-m telescope of the MPIfR (Max-Planck-Institut für Radioastronomie) at Effelsberg.
\end{acknowledgements}

%

   \bibliographystyle{aa} 
   \bibliography{mybib.bib} 

\end{document}